\documentclass[aoas,MSNbibl,nameyear,dvips]{arximspdf}
\usepackage{graphicx}

\doi{10.1214/11-AOAS487}
\volume{5}
\issue{4}
\pubyear{2011}
\firstpage{2549}
\lastpage{2571}

\begin{document}
\begin{frontmatter}

\title{Residual analysis methods for space--time point processes with
applications to earthquake forecast models in California\thanksref{T1}}
\runtitle{Residual analysis for earthquake forecast models}

\thankstext{T1}{Supported by the Southern California
Earthquake Center. SCEC is funded by NSF Cooperative Agreement
EAR-0106924 and USGS Cooperative Agreement 02HQAG0008. The SCEC
contribution number for this paper is 1495.}

\begin{aug}
\author[A]{\fnms{Robert Alan} \snm{Clements}\corref{}\ead[label=e1]{clements@stat.ucla.edu}},
\author[A]{\fnms{Frederic Paik} \snm{Schoenberg}\ead[label=e2]{frederic@stat.ucla.edu}}\\
and
\author[B]{\fnms{Danijel} \snm{Schorlemmer}\ead[label=e3]{ds@usc.edu}\ead[label=e4]{ds@gfz-potsdam.de}}
\runauthor{R. A. Clements, F. P. Schoenberg and D. Schorlemmer}
\affiliation{University of California, Los Angeles,
University of California, Los Angeles, and University of
Southern California and GFZ Potsdam}
\address[A]{R. A. Clements\\
F. P. Schoenberg\\
Department of Statistics \\
University of California\\
8125 Math Sciences Building \\
Los Angeles, California 90095-1554 \\
USA\\
\printead{e1}\\
\hphantom{E-mail: }\printead*{e2}} 
\address[B]{D. Schorlemmer\\
Southern California Earthquake Center \\
University of Southern California \\
3651 Trousdale Parkway \\
Los Angeles, California 90089-0740 \\
USA\\
\printead{e3}\\
and\\
German Research Center for Geosciences \\
Telegrafenberg \\
14473 Potsdam \\
Germany \\
\printead{e4}}
\end{aug}

\received{\smonth{9} \syear{2010}}
\revised{\smonth{5} \syear{2011}}

%
\begin{abstract}
Modern, powerful techniques for the residual analysis of
spatial-temporal point process models are reviewed and compared. These
methods are applied to California earthquake forecast models used in
the Collaboratory for the Study of Earthquake Predictability (CSEP).
Assessments of these earthquake forecasting models have previously been
performed using simple, low-power means such as the L-test and N-test.
We instead propose residual methods based on rescaling, thinning,
superposition, weighted \mbox{K-functions} and deviance residuals. Rescaled
residuals can be useful for assessing the overall fit of a model, but
as with thinning and superposition, rescaling is generally impractical
when the conditional intensity $\lambda$ is volatile. While residual
thinning and superposition may be useful for identifying spatial
locations where a model fits poorly, these methods have limited power
when the modeled conditional intensity assumes extremely low or high
values somewhere in the observation region, and this is commonly the
case for earthquake forecasting models. A recently proposed hybrid
method of thinning and superposition, called super-thinning, is a more
powerful alternative. The weighted K-function is powerful for
evaluating the degree of clustering or inhibition in a~model. Competing
models are also compared using pixel-based approaches, such as Pearson
residuals and deviance residuals. The different residual analysis
techniques are demonstrated using the CSEP models and are used to
highlight certain deficiencies in the models, such as the
overprediction of seismicity in inter-fault zones for the model
proposed by Helmstetter, Kagan and
Jackson [\textit{Seismological Research Letters} \textbf{78} (2007)
78--86], the underprediction of the model proposed by Kagan, Jackson
and Rong [\textit{Seismological Research Letters} \textbf{78} (2007)
94--98] in forecasting seismicity around the Imperial, Laguna Salada,
and Panamint clusters, and the underprediction of the model proposed by
Shen, Jackson and Kagan [\textit{Seismological Research Letters}
\textbf{78} (2007) 116--120] in forecasting seismicity around the
Laguna Salada, Baja, and Panamint clusters.
\end{abstract}

%

\end{frontmatter}

\section{Introduction}

Recent statistical developments in the assessment of space--time point
process models have resulted in new, powerful model evaluation tools.
These tools include residual point process methods such as thinning,
superposition and rescaling, comparative quadrat methods such as
Pearson residuals and deviance residuals, and weighted second-order
statistics for assessing particular features of a model such as its
background rate or the degree of spatial clustering.

Unfortunately, these methods have not yet become widely used in
seismology. Indeed, recent efforts to assess and compare different
space--time models for earthquake occurrences have led to developments
such as the Regional Earthquake Likelihood Models (RELM) project
[\citet{Field2007}] and its successor, the Collaboratory for the Study
of Earthquake Predictability (CSEP) [\citet{Jordan2006}]. The RELM
project was initiated to create a~variety of earthquake forecast models
for seismic hazard assessment in California. Unlike previous projects
that were addressing earthquake forecast modeling for seismic hazard
assessment, the RELM participants decided to develop a~multitude of
competing forecasting models and to rigorously and \textit{prospectively}
test their performance in a dedicated testing center [\citet{Schor}].
With the end of the RELM project, the forecast models became available
and the development of the testing center was done within the scope of
CSEP. CSEP inherited not only all models developed for RELM and is
testing them for the previously defined period of 5 years, but also a
suite of forecast performance tests that was developed during the RELM
project. In RELM, a community consensus was reached that all models
will be tested with these tests [\citet{Jackson}, \citet
{CSEP}]. The tests
include the Number or N-Test that compares the total forecasted rate
with the observation, the Likelihood or L-Test that assesses the
quality of a forecast in the likelihood space, and the Likelihood-Ratio
or R-Test that compares the performance of two forecast models.
However, over time several drawbacks of these tests were discovered
[\citet{Schorlemmer2010}] and the need for more and powerful tests
became clear to better discern between closely competing models. The
N-test and L-test simply compare the quantiles of the total numbers of
events in each bin or likelihood within each bin to those expected
under the given model, and the resulting low-power tests are typically
unable to discern significant lack of fit unless the overall rate of
the model fits extremely poorly. Further, even when the tests do reject
a model, they do not typically indicate \textit{where} or \textit{when} the
model fits poorly, or how it could be improved.

The purpose of the current paper is to review modern model evaluation
techniques for space--time point processes and to demonstrate their use
and practicality on earthquake forecasting models for California. The
RELM project represents an ideal test case for this purpose, as a
variety of relevant, competing space--time models are included, and
these models yield genuinely prospective forecasts of earthquake rates
based solely on prior data. The rates are specified per bins which are
spatial-magnitude-temporal volumes (called pixels in the statistical
domain). These bins have been predefined in a community consensus
process in order to have the model forecast rates in the exact same
bins. The models' forecasts translate into strongly different estimates
of seismic hazard. Its accurate estimation is important for seismic
hazard assessment, urban planning, disaster preparation efforts and in
the pricing of earthquake insurance premiums
[\citet{Bolt}], so distinguishing among competing models is an
extremely important task.

In Section \ref{sec2} we describe a group of earthquake forecast models to be
evaluated, along with the observed earthquake occurrences used to
assess the fit of the models. The methods currently used by
seismologists for model evaluation are briefly reviewed in Section \ref{sec3}.
Pixel-based residuals for model comparison are discussed in Section \ref{sec4}.
In Section \ref{sec5} weighted second-order statistics, primarily the weighted
K-function, are investigated. Section \ref{sec6} reviews various residual
methods based on rescaling, thinning and superposition, and
introduces and applies the method of
super-thinning. Section \ref{sec7}
summarizes some of the benefits and weaknesses of these tools.

\section{CSEP earthquake forecast models and earthquake occurrence
catalogs}\label{sec2}

CSEP expanded and now collects and evaluates space--time earthquake
forecasts for different regions around the world, including California,
Japan, New Zealand, Italy, the Northwest Pacific, the Southwest Pacific
and the entire globe. The forecasts are evaluated in testing centers in
Japan, Switzerland, New Zealand and the United States. The U.S.
testing center is located at the Southern California Earthquake Center
(SCEC) and hosts forecast experiments for California, the Northwest and
Southwest Pacific, and the global experiments. We have chosen to apply
a variety of measures to assess the fit of a collection of the
California forecast models currently being tested at SCEC.

The forecast models are arranged in classes according to their forecast
time period: five-year, three-month and one-day. There are two types of
forecasts, rate-based and alarm-based. Within the five-year group are a
set of rate-based models developed as part of the RELM project. In this
paper we evaluate the RELM project rate-based one-day and five-year
models, and will be ignoring the three-month models due to their very
recent introduction to the CSEP testing center.

All CSEP forecasts are grid-based, providing a forecast in each
spatial-magnitude bin within a given time window. For the one-day
models, each bin is of size $0.1^{\circ}$ longitude (lon) by $0.1^{\circ
}$ latitude (lat) by $0.1$ units magnitude for earthquake magnitudes
ranging from 3.95 to 8.95. For magnitudes 8.95--10, there is a single
bin of size $0.1^{\circ}$ by $0.1^{\circ}$ by $1.05$ units of
magnitude. The RELM forecasts are identical, except with a lower
magnitude bound of 4.95 instead of 3.95. For each bin, an expected
number of earthquakes in the forecast period is forecasted.

There are five models in the RELM project that are considered
main\-shock$+$ aftershock models. These models forecast both mainshocks and
aftershocks with a single forecast for a period of five years. Models
proposed in \citet{Helmstetter} and \citet{Kaganforecast},
which we
will call models A and B, respectively, base their forecasts
exclusively on previous seismicity. The model proposed in \citet{Shen},
denoted model C here, is based on other geodetic or geological data.
All RELM models are five-year forecasts, beginning 1 January 2006,
00:00 UTC and ending 1 January 2011, 00:00 UTC. CSEP is also testing
two one-day forecast models: The Epidemic-Type Aftershock Sequences
(ETAS) model [\citet{Zhuang2004}, \citet{Ogata2006}] and the
Short-Term Earthquake Probabilities (STEP) model
[\citet{Gerstenberger2005}] since September of 2007. Both of these
models produce forecasts based exclusively on prior seismicity.

CSEP evaluates the RELM models using a lower magnitude cutoff of~4.95.
Because there are so few earthquakes of magnitude~4.95 and higher in
the catalog over the observed period we use a lower magnitude cutoff of~3.95 instead. The forecasts for models A, B and C were extrapolated
using each model's fitted magnitude distribution. Models A and B assume
the magnitude distribution follows a tapered Gutenberg--Richter law
[\citet{gr}] with a \textit{b}-value of 0.95 and a corner magnitude
of 8.0.
Model C uses a \textit{b}-value of 0.975 and the same corner magnitude.
Model A adjusts the magnitude distribution in a small region in
northern California influenced by geothermal activity
(122.9$^{\circ}$W${}<{}$lon${}<{}$122.7$^{\circ}$W and
$38.7^{\circ}$N${}<{}$lat${}<{}$38.9$^{\circ}$N)
by using a \textit{b}-value of 1.94 instead of 0.95.

\begin{figure}

\includegraphics{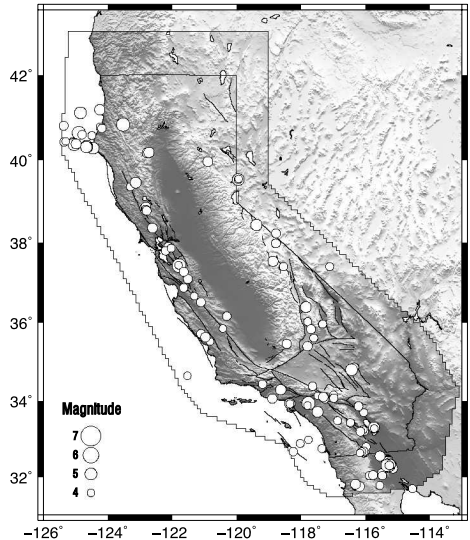}

\caption{Locations of earthquakes with magnitude $M\geq3.95$ in the
RELM testing region.}\label{alleqs}
\end{figure}

Earthquake catalogs containing the estimated earthquake hypocenter
locations and magnitudes were obtained from the Advanced National
Seismic System (ANSS). From 1 January 2006 to 1 September 2009 there
were 142 shallow earthquakes with a magnitude of 3.95 or larger which
occurred in RELM's spatial-temporal window (see Figure \ref{alleqs}).
%
Note that each RELM model does not necessarily produce a forecasted
seismicity rate for every pixel in the space--time region. Hence, each
model essentially has its own relevant spatial-temporal observation
region, and thus we\vadjust{\goodbreak} may have different numbers of observed earthquakes
corresponding to different models. For instance, all 142 recorded
earthquakes from 1 January 2006 to 1 September 2009 corresponded to
pixels where model A made forecasts, but only 81 corresponded to pixels
where model B made forecasts, and 86 where model C made forecasts. 85
earthquakes of magnitude 3.95 or greater occurred since 1~September of
2007, all of which corresponded to forecasts made by ETAS but only 83
of which corresponded to forecasts made by STEP.

\section{L-test and N-test}\label{sec3}

CSEP initially implemented two numerical summary tests, called the
Likelihood-test (L-test) and the Number-test (N-test), to evaluate the
fit of the earthquake forecast models they collect. A full description
of these methods can be found in \citet{CSEP}. These goodness-of-fit
tests are similar to other numerical goodness-of-fit summaries such as
the Akaike Information Criterion [\citet{AIC}] and the Bayesian
Information Criterion [\citet{BIC}] in that they provide a score
for the
overall fit of the model without indicating where the model may be
fitting poorly.

The L-test, described in \citet{CSEP}, works by first simulating
some fixed number $s$ of realizations from the forecast model. The
log-likelihood ($\ell$) is computed for the observed earthquake catalog
($\ell_{\mathrm{obs}}$) and each simulation ($\ell_{j}$, for $j=1,2,\ldots
,s$). The quantile score, $\gamma$, is defined as the fraction of
simulated likelihoods that are less than the observed catalog likelihood:
\[
\gamma= \frac{\sum _{j=1}^s \mathbf{1}_{\{\ell_j < \ell_{\mathrm
{obs}}\}}}{s},
\]
where ${\mathbf 1}$ denotes the indicator function.
If $\gamma$ is close to zero, then the model is considered to be
inconsistent with the data, and can be rejected. Otherwise, the model
is not rejected and further tests are necessary.

The N-test is similar to the L-test, except that the quantile score
examined is instead the fraction of simulations that contain fewer
points than the actual observed number of points in the catalog,
$N_{\mathrm{obs}}$. That is,
\[
\delta= \frac{\sum _{j=1}^s {\mathbf 1}_{\{N_j < N_{\mathrm{obs}}\}
}}{s},
\]
where $N_j$ is the number of points in the $j$th simulation of the
model. With the N-test, the model is rejected if $\delta$ is close to
$0$ or $1$.
If a model is underpredicting or overpredicting the total number of
earthquakes, then $\delta\sim1$ or~$0$, respectively, and the model
will likely be rejected with the N-test.

Table \ref{taB} shows results for the L- and N-test for selected
models. The L-test would lead to rejection of models A, B, C and STEP
as seen by the very low $\gamma$ scores. The ETAS model would not be
rejected based on the $\gamma$ score alone, requiring the application
of the N-test for a final decision. At the $5\%$ level of significance,
the $\delta$ scores indicate that the STEP model is underpredicting the
total number of earthquakes, while models A, B, C and ETAS are
significantly overpredicting earthquake rates.

\begin{table}
\caption{Results of the L and N-test. Listed
are the observed log-likelihoods, $\ell_{\mathrm{obs}}$, the L-test
$\gamma$ scores, the observed number of events, $N_{\mathrm{obs}}$ and
the N-test $\delta$ scores. $\delta$ scores that are bold-faced are
significant at the $5\%$ level leading to rejection of the forecast}
\label{taB}
\begin{tabular*}{\tablewidth}{@{\extracolsep{\fill}}lrccc@{}}
\hline
\textbf{Model} & \multicolumn{1}{c}{$\bolds{\ell_{\mathrm{obs}}}$}
& \multicolumn{1}{c}{$\bolds{\gamma}$} & \multicolumn{1}{c}{$\bolds{N_{\mathrm{obs}}}$}
& \multicolumn{1}{c@{}}{$\bolds{\delta}$} \\
\hline
\multicolumn{5}{@{}c@{}}{Mainshock$+$Aftershock} \\
[4pt]
A. Helmstetter & $-$22881.46 & 0.000 & 142 & \textbf{0.000} \\
B. Kagan & $-$10765.43 & 0.008 & \hphantom{0}81 & \textbf{0.001} \\
C. Shen & $-$10265.20 & 0.002 & \hphantom{0}86 & \textbf{0.043} \\
[4pt]
\multicolumn{5}{@{}c@{}}{Daily}\\
[4pt]
ETAS & $-$387.69 & 1.00\hphantom{0} & \hphantom{0}85 & \textbf{0.00}\hphantom{\textbf{0}} \\
STEP & $-$50.43 & 0.00\hphantom{0} & \hphantom{0}83 & \textbf{0.99}\hphantom{\textbf{0}} \\
\hline
\end{tabular*}
\end{table}

Unfortunately, in practice, both statistics $\gamma$ and $\delta$ test
essentially the same thing, namely, the agreement between the observed
and modeled \textit{total} number of points. Indeed, for a typical model,
the likelihood for a given simulated earthquake catalog depends
critically on the number of points in the simulation.

\section{Pixel-based methods}\label{sec4}

Baddeley et~al. (\citeyear{Baddeley}) introduced methods for residual analysis of
purely spatial point processes, based on comparing the total number
of\vadjust{\goodbreak}
points within predetermined bins to the number forecast by the model.
Such methods extend readily to the spatial-temporal case, and are quite
natural for evaluating the CSEP forecasts since the models are
constrained to have a constant conditional intensity within
prespecified bins. The differences between observed and expected
numbers of events within bins can be standardized in various ways, as
described in what follows.

\subsection{Preliminaries}\label{sec41}

Earthquake occurrence times and locations are typically modeled as
space--time point processes, with the estimated epicenter or hypocenter
of each earthquake representing its spatial location. Along with each
observation, one may also record several \textit{marks} which may be used
in the model to help forecast future events; an important example of a
mark is the magnitude of the event. Space--time point process models are
often characterized by their associated conditional intensity,
$\lambda(t, \mathbf{x})$, that is, the infinitesimal rate at which one
expects points to occur around time $t$ and location ${\mathbf x}$, given
full information on the occurrences of points prior to time~$t$, and
given the marks and possibly other covariate information observed
before time $t$. Note that due to the lack of a natural ordering of
points in the plane, purely spatial point processes are typically
characterized by their Papangelou intensities [\citet{Pap2}],
which may
be thought of as the limiting rate at which points are expected to
accumulate within balls centered at location ${\mathbf x}$ given what \textit{other} points have occurred at all locations outside of these balls, as
the size of the balls shrink to zero. For a review of point processes
and conditional intensities, see \citet{VJ}.

An aggregate conditional intensity is derived for each spatial bin for
all models by summing the forecast rates over all magnitude bins and
then dividing the sum by the area of each pixel. Since we are
evaluating the five-year models A, B and C after only 44 of the 60
months of the forecast period have elapsed, their conditional
intensities are scaled by a factor of ${44}/{60}$.

\subsection{Raw and Pearson residuals}\label{sec42}

Consider a model $\hat\lambda(t,x,y)$ for the conditional intensity at
any time $t$ and location $(x,y)$. \textit{Raw residuals} may be defined
following \citet{Baddeley} as simply the number of observed points
minus the number of expected points in each pixel, that is,
%
\begin{equation}\label{rawres}
R(B_{i}) = N(B_{i}) - \int_{B_{i}}\hat{\lambda}(t,x,y)\,dt\,dx\,dy,
\end{equation}
where $N(B_{i})$ is the number of points in bin $i$.
Note that \citet{Baddeley} consider only the case of purely
spatial point processes characterized by their Papangelou intensities;
\citet{Zhuang2006} showed that one may nevertheless extend the
definition to the spatial-temporal case using the conventional
conditional intensity as in (\ref{rawres}).

One may wish to rescale the raw residuals in such a way that they have
mean 0 and variance\vadjust{\goodbreak} approximately equal to 1. The \textit{Pearson
residuals} are defined as
\[
R_{\mathrm{P}}(B_{i}) = \sum_{(t_{j},x_{j},y_{j}) \in B_{i}}\frac
{1}{\sqrt{\hat{\lambda}(t_{j},x_{j},y_{j})}}-\int_{B_{i}}\sqrt{\hat
{\lambda}(t,x,y)}\,dt \,dx \,dy
\]
for all $\hat{\lambda}(t_{i},x_{i},y_{i})>0$. These are analogous to
the Pearson residuals in Poisson log-linear regression.

\begin{figure}

\includegraphics{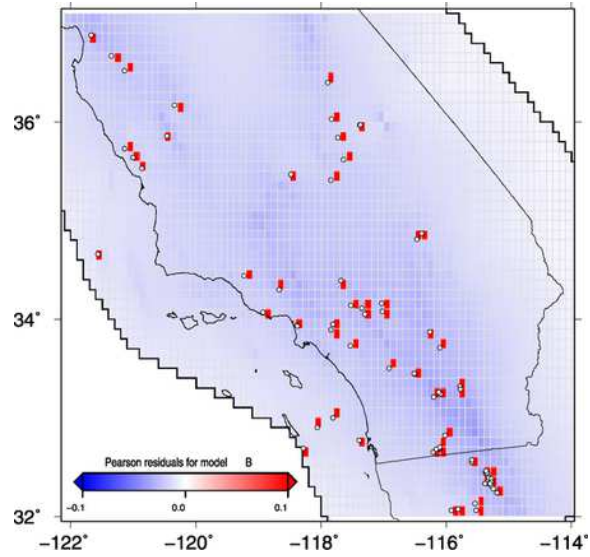}

\caption{Pearson residuals for model B. The maximum observed Pearson
residual is 2.817.}\label{Kaganpearson}
\end{figure}

Both STEP and model C have several pixels with forecasted conditional
intensities of 0, which complicates the standardization of the
corresponding residuals for these two models. Pearson residuals were
obtained for each of the remaining models. For instance, Figure
\ref{Kaganpearson} shows that the largest Pearson residual for model B
is 2.817 located in a pixel in Mexico, just south of the California
border near the Imperial Valley fault zone ($\mathrm{lon}\approx
115.3^{\circ}$W and $\mathrm{lat} \approx32.4^{\circ}$N), which is the location
of a large cluster of earthquakes. Another very large residual for
model B can be seen just above the San Bernardino and Inyo county
border near the Panamint Valley fault zone ($\mathrm{lon} \approx
117.0^{\circ}$W and $\mathrm{lat} \approx36.0^{\circ}$N). This is also the
location of the largest ETAS Pearson residual (2.221). The largest
Pearson residual for model A (4.068) is located at a small earthquake
cluster near the Peterson Mountain fault northwest of Reno, Nevada
($\mathrm{lon}\approx199.9^{\circ}$W and $\mathrm{lat}\approx39.5^{\circ}$N).

%

Note that when spatial-temporal bins are very small and/or the
estimated conditional intensity in some bins is very low, as in this
example, the raw and especially the standardized residuals are highly
skewed. In such cases, the residuals in such pixels where points happen
to occur tend to dominate, and the skew may complicate the analysis.
Indeed, Pearson residuals fail to provide much useful information about
the model's fit in the other pixels where earthquakes did not happen to
occur, and graphical displays of the Pearson residuals tend to
highlight little more than the locations of the earthquakes themselves.
Therefore, while Pearson and raw residuals may help to identify
individual bins containing earthquakes that require an adjustment in
their forecasted rates, Pearson and raw residuals generally fail to
identify other locations where the models may fit relatively well or poorly.

\subsection{Deviance residuals}\label{sec43}

A useful method for comparing models is using the deviance residuals
proposed by \citet{Wong10}, in analogy with deviances defined for
generalized linear models in the regression framework. As with Pearson
residuals, $S$ is divided into evenly spaced bins, and the differences
between the log-likelihoods within each bin for the two competing
models are examined. Given two models for the conditional intensity,
$\hat\lambda_{1}$ and $\hat\lambda_{2}$, the deviance residual in
each bin, $B_{i}$, of $\hat{\lambda}_{1}$ against $\hat{\lambda}_{2}$
is given by
\begin{eqnarray*}
R_{\mathrm{D}}(B_{i}) &=& \sum_{i\dvtx(t_{i},x_{i},y_{i})\in B_{i}} \log{(\hat
{\lambda}_{1}(t_{i},x_{i}, y_{i}))} - \int_{B_{i}} \hat{\lambda
}_{1}(t,x,y)\,dt \,dx \,dy \\
&&{}- \biggl(\sum_{i\dvtx(t_{i},x_{i},y_{i})\in B_{i}}\log{(\hat{\lambda
}_{2}(t_{i},x_{i},y_{i}))} - \int_{B_{i}} \hat{\lambda}_{2}(t,x,y)\,dt \,dx
\,dy \biggr).
\end{eqnarray*}

Positive residuals imply that the model $\hat{\lambda}_{1}$ fits better
in the given pixel and negative residuals imply that $\hat{\lambda
}_{2}$ provides better fit. By simply taking the sum of the deviance
residuals, $\sum_{{i}}R_{\mathrm{D}}(B_{i})$, we obtain a log-likelihood
ratio score, giving us an overall impression of the improvement in fit
from the better fitting model.
If $\hat{\lambda}_{1}$ or $\hat{\lambda}_{2}$ is estimated, then one
may use this estimate in computing the deviance residuals, and
similarly if $\hat{\lambda}_{1}$ or $\hat{\lambda}_{2}$ is given, that
is, not estimated, then one would simply use this given model in
computing the residuals.

\begin{figure}

\includegraphics{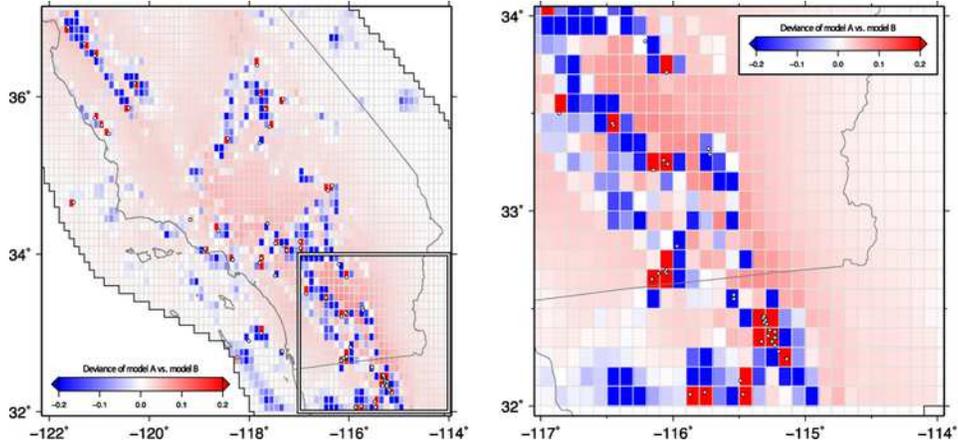}

\caption{Left panel \textup{(a)}: deviance residuals for model A versus B. Sum
of deviance residuals is 84.393. Right panel \textup{(b)}: close-up of deviance
residuals for model A versus B near the Imperial fault.}\label{devcombined}
\end{figure}

Figure \ref{devcombined}(a) shows the deviance residuals for model A
versus model B. Model A outperforms model B in almost all locations
where earthquakes actually occurred, and, in particular, model A
forecasts the Imperial earthquake cluster and another cluster near the
Laguna Salada and Yuha Wells faults just north of the California--Mexico
border ($\mathrm{lon} \approx116.0^{\circ}$W and $\mathrm{lat} \approx32.7^{\circ}$N)
much better than model B. The pixel with the largest residual,
highlighted in Figure \ref{devcombined}(b), is located in the Imperial
cluster. Model B seems to fit better in several selected areas, mostly
regions close to known faults but where earthquakes did not happen to
occur in the time span considered. In most locations, however,
including the vast majority of locations far from seismicity, model A
offers better fit, as model B tends to overpredict events in these
locations more than model A. Overall, the log-likelihood ratio score is
84.393, indicating a significant improvement from model A compared to
model B.

%
%

%
\begin{figure}[b]

\includegraphics{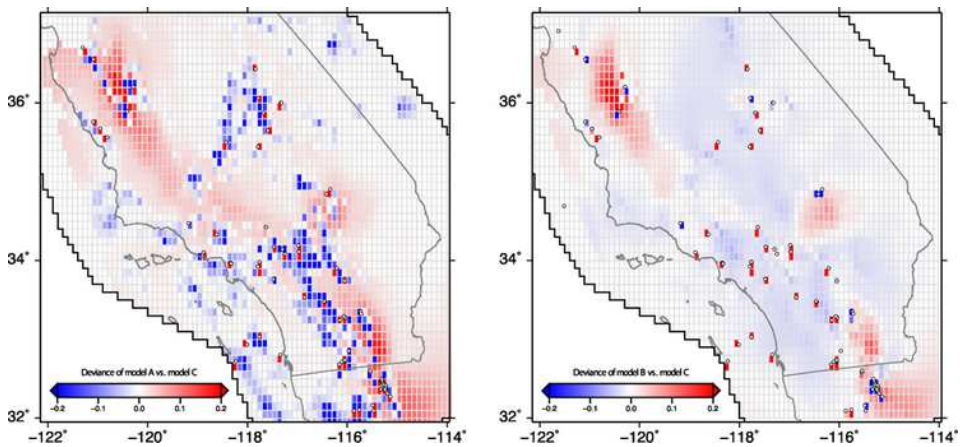}

\caption{Left panel \textup{(a)}: deviance residuals for model A versus C. Sum
of deviance residuals is 86.427. Right panel \textup{(b)}: deviance residuals
for model B versus C. Sum of deviance residuals is
$-$7.468.}\label{devcombined2}
\end{figure}

Results are largely similar for model A versus model C, as seen in
Figure~\ref{devcombined2}(a), with model A forecasting the rate at all
observed earthquake clusters, including a cluster at the extreme
southern end of the observation region on the Baja, Mexico peninsula
($\mathrm{lon}\approx116.3$W and $\mathrm{lat}\approx31.8$N), more accurately than
model C. Overall, model A offers substantial improvement over model C
with a likelihood ratio score of 86.427. Residuals for model B versus
model C can be seen in Figure \ref{devcombined2}(b). Model C forecasts
the rate near the Imperial cluster better, and model B forecasts more
accurately around the Laguna Salada cluster. There are vast regions
where model B outperforms model C and vice versa. Overall, model C fits
slightly better than model B, with a likelihood ratio score of $-$7.468.
Deviance residuals for ETAS versus STEP (not shown) reveal that the
ETAS model performs somewhat better for this data set overall, with a
log-likelihood ratio score of 76.261, providing substantially more
accurate forecasts in nearly all locations, especially where
earthquakes occur.

\section{Weighted second-order statistics}\label{sec5}

A common model assessment tool used for detecting clustering or
inhibition in a point process is Ripley's K-function [\citet
{Ripley}], defined as the average number of points within $r$ of any
given point divided by the overall rate $\lambda$, and is typically
estimated via
\[
\hat{\mathrm{K}}(r) = AN^{-2}\sum_{ i<j,
\|x_{i}-x_{j}\|<r}s(\mathbf{x}_{i}, \mathbf{x}_{j}),
\]
where $A$ is the area of the observation region, $N$ is the total
number of observed points, and $s(\mathbf{x}_{i}, \mathbf{x}_{j})^{-1}$
is the proportion of area of the ball centered at $\mathbf{x}_{i}$ and
passing through $\mathbf{x}_{j}$ that falls within the observation
region [see \citet{Ripley}, \citet{Cressie}]. For a
homogeneous Poisson process
in ${\mathbf R}^2$, K$(r)=\pi r^{2}$, \citet{Besag} suggested a variance
stabilized version of the K-function, called the L-function, given by
L$(r)=\sqrt{{\mathrm{K}(r)}/{\pi}}$.

The null hypothesis for most second-order tests such as Ripley's
K-function is that the point process is a homogeneous Poisson process.
\citet{Stark} argues that this is a poor null hypothesis for the
case of earthquake occurrences because a homogeneous Poisson model fits
so poorly to actual data. \citet{Adelfio09} described a variety of
weighted analogues of second-order tests that are useful when the null
hypothesis in question is more general. Most useful among these is the
weighted analogue of Ripley's K-function, first introduced by
\citet{Baddeley2}. They discussed the case where the null
model~$\hat{\lambda}_{0}$, can be any inhomogeneous Poisson process, and this
was extended by \citet{Veen} to the case of non-Poisson processes as
well. The weighted K-function is useful for testing the degree of
clustering in the model, and was used by \citet{Veen} to assess a
spatial point process model fitted to Southern California earthquake
data. The standard estimate of the weighted K-function is given by
\[
\mathrm{K}_{\mathrm{W}}(r) = \frac{b}{\int_S {\hat{\lambda}_{0}(\mathbf
{x})\,d\mathbf{x}}} \sum_{i}\hat{\lambda}_{0}(\mathbf{x}_{i})^{-1}\sum
_{j\neq i}\hat{\lambda}_{0}(\mathbf{x}_{j})^{-1}{\mathbf 1}_{\{|\mathbf
{x}_{j}-\mathbf{x}_{i}|\leq r\}},
\]
where $b=$min($\hat{\lambda}$), ${\mathbf 1}$ is the indicator function,
and $\hat{\lambda}_{0}(\mathbf{x}_{i})$ is the conditional intensity at
point $\mathbf{x}_{i}$ under the null hypothesis. Edge-corrected
modifications can also be used, especially when the observed space is irregular.
\citet{guan} proposed a local empirical K-function which can
assess lack-of-fit in subsets of $S$ and can be compared to the
weighted K-function applied globally to $S$. Here, we apply the
weighted K-function globally to derive an overall impression of each
model's lack of fit.

As with Ripley's K-function, under the null hypothesis, for a spatial
point process with intensity $\lambda_0$, $\mathrm{K}_{\mathrm{W}}(r)=\pi
r^{2}$ [\citet{Veen}]. To obtain a centered and standardized
version, one can also transform the weighted K-function into a weighted
L-function as before, and plot $\mathrm{L}_{\mathrm{W}}(r)-r = \sqrt
{\mathrm{K}_{\mathrm{W}}(r)/ \pi} - r$ versus~$r$.

%
\begin{figure}

\includegraphics{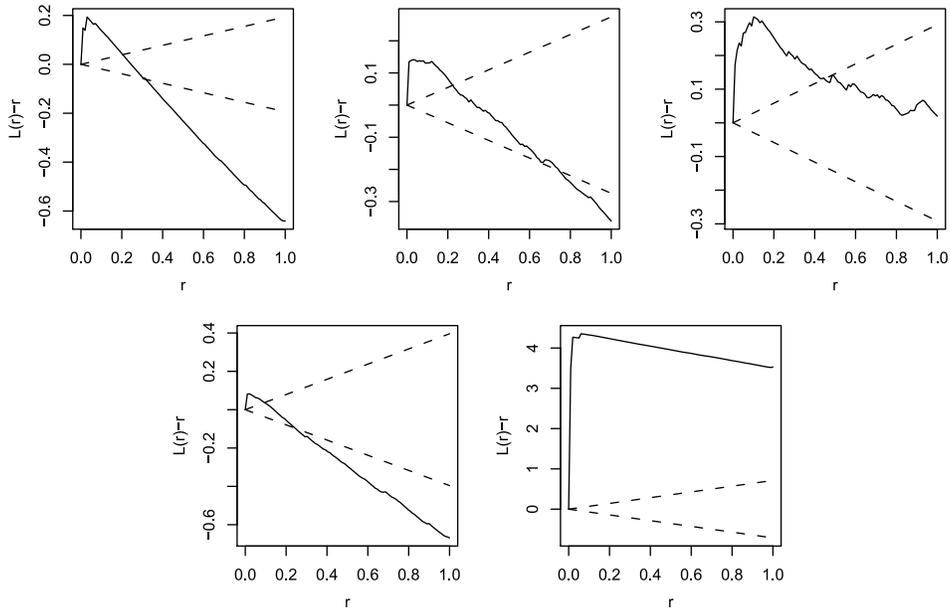}

\caption{Estimated centered weighted L-function (solid curve) and 95\%
confidence bands (dashed curves). Top-left panel: \textup{(a)} model A.
Top-center panel: \textup{(b)} model B. Top-right panel:
\textup{(c)} model C.
Bottom-left panel: \textup{(d)} ETAS.
Bottom-right panel: \textup{(e)} STEP.}\label{allwk}
\end{figure}

Space--time versions of the L-function have been proposed, but for the
purpose of examining, in particular, the range and degree of purely
spatial clustering in each model, it seems preferable to apply the
purely spatial weighted L-function previously described, after first
integrating the conditional intensities of the ETAS and STEP models
over time. Figure \ref{allwk} shows the estimated centered weighted
L-functions for the five models considered here, along with 95\%
confidence bounds based on the normal approximation in \citet{Veen},
who showed that asymptotically, the distribution of the weighted
K-function should generally obey
%
\begin{equation}\label{wkbounds}
\mathrm{K}_{\mathrm{W}}(r) \sim N\biggl(\pi r^2, \frac{2\pi r^2 A}{[\int
_S {\hat{\lambda}_{0}(\mathbf{x})\,d\mathbf{x}}]^2}\biggr).
\end{equation}
The catalog of observed earthquakes is significantly more clustered
than would be expected according to model A, especially within
distances of~$0.2$ degrees of longitude/latitude, or approximately $22.2$ km.
However, at distances greater than~$0.3^{\circ}$, or approximately
$33.3$ km, the observed data exhibit greater inhibition than one would
expect according to model A. This suggests that model A is
underpredicting the degree of clustering in the observed seismicity and
may be generally underpredicting the seismicity rate within highly
active seismic areas, and may be overpredicting seismicity elsewhere.
Results are similar for model B and the ETAS model. The estimated
L-function for model C shows significantly more clustering of the
(weighted) seismicity than one would expect within distances of
$0.4^{\circ}$ or $44.4$ km, that is, model C is significantly
underpredicting the degree of clustering within this range, but seems
consistent with the data outside of this range. The estimated
L-function shows clear discrepancies between the STEP model and the
data, as the (weighted) seismicity is significantly more clustered than
one would expect according to the model at both small and large
distances. These results are not surprising
considering that STEP tends to underpredict seismicity overall:
according to the STEP forecasts, one would expect only 63 earthquakes
in total during the period in which 85 occurred. By contrast,
ETAS tends to overpredict the overall rate, forecasting more than 114
earthquakes in this same period.

\section{Residual point process methods}\label{sec6}

As shown in Section \ref{sec42}, when the spatial-temporal pixels are
small, the distribution of raw and Pearson residuals tend to be highly
skewed, and this limits their utility. When pixels are larger, however,
a~drawback of pixel-based residuals is that considerable information is
lost in aggregating over the pixels. Instead, one may wish to examine
the extent to which the data and model agree, without relying on such
aggregation. One way to perform such an assessment is to transform the
points of the process, by rescaling, thinning, superposition or
superthinning, to form a new point process that should be a homogeneous
Poisson process if and only if the model used to govern this
transformation is correct. The residual points can then be assessed for
inhomogeneity as a means of evaluating the goodness of fit of the
underlying model.

\subsection{Rescaled residuals}

\citet{Meyer} observed that the temporal coordinates of a
multivariate point process can be rescaled according to the integrated
conditional intensity in order to form a sequence of stationary Poisson
processes. For a space--time point process, one may thus rescale one
axis, for example, the $x$-axis, moving each observation $(t_i,x_{i},
y_{i})$ to the new rescaled position $(t_i,
\int_{0}^{x_{i}}\hat{\lambda}(t,x,y) \,dx, y_{i} )$, and assess the
space--time homogeneity of the resulting process. This sort of method
was used by \citet{Ogata} for model evaluation for the purely temporal
case and by \citet{Rick} for the spatial-temporal case. The spatial
homogeneity of these residual points may be assessed, for instance
using Ripley's K-function.

\begin{figure}

\includegraphics{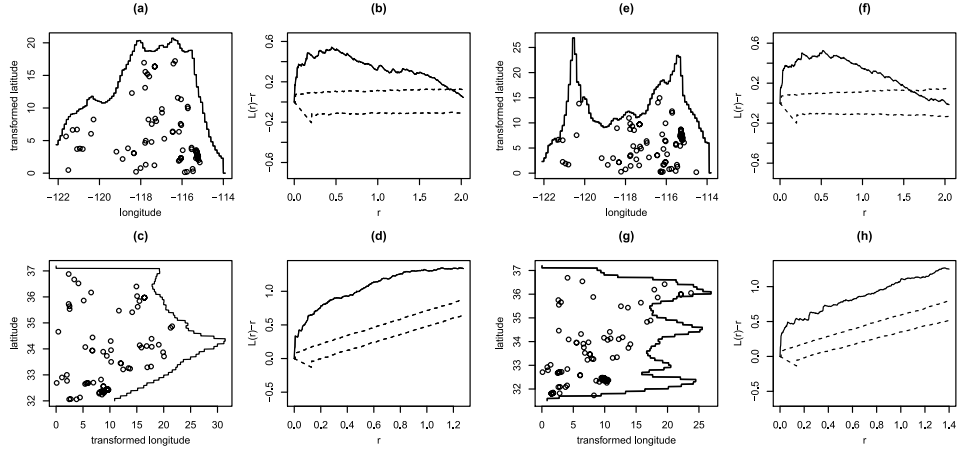}

\caption{Rescaled residuals and transformed space for models B and C.
\textup{(a)}: vertically rescaled residuals for model B. \textup{(b)}:
estimated centered
L-function for vertically rescaled residuals (solid line) and middle
95\% ranges of estimated centered L-functions for 1\mbox{,}000 simulated
homogeneous Poisson processes (dashed lines). \textup{(c)}: horizontally
rescaled residuals for model B. \textup{(d)}: estimated centered L-function for
horizontally rescaled residuals (solid line) and middle 95\% ranges of
estimated centered L-functions for 1\mbox{,}000 simulated homogeneous Poisson
processes (dashed lines). \textup{(e)}: vertically rescaled residuals for model
C. \textup{(f)}:~estimated centered L-function for vertically rescaled residuals
(solid line) and middle 95\% ranges of estimated centered L-functions
for 1\mbox{,}000 simulated homogeneous Poisson processes (dashed lines). \textup{(g)}:
horizontally rescaled residuals for model B. \textup{(h)}: estimated centered
L-function for horizontally rescaled residuals (solid line) and middle
95\% ranges of estimated centered L-functions for 1\mbox{,}000 simulated
homogeneous Poisson processes (dashed lines).}\label{combresc}
\end{figure}

If $\lambda$ is spatially volatile, the transformed space bounding the
rescaled residuals can be highly irregular, which makes it difficult to
detect uniformity using the K-function. In this case, one can rescale
the points along a~different axis as in \citet{Rick2} and see if there
is any improvement. Unfortunately, most CSEP forecast models have
volatile conditional intensities, resulting in a highly irregular
boundary regardless of which axis is chosen for rescaling. In such
cases, the K-function is dominated by boundary effects and has little
power to detect excessive clustering or inhibition in the residuals.
Figure \ref{combresc} shows the rescaled residuals for models B and~C,
%
which had the most well behaved of the rescaled residuals for the five
models we considered. There is significant clustering in both the
vertically and horizontally rescaled residuals for all five models,
apparently due to clustering in the observations not adequately
accounted for by the models, the most noticeable of which is the very
large Imperial cluster. One must be somewhat cautious, however, in
interpreting rescaled residuals, because patterns observed in the
points in the rescaled coordinates may be difficult to interpret.

\subsection{Thinned residuals}

Thinned residuals are a modification to the simulation techniques used
by \citet{Lewis} and \citet{Ogata81}, and, as shown in
\citet{Rick},
are useful for assessing the spatial fit of a~space--time point process
model and revealing locations where the model is fitting poorly. Unlike
rescaled residuals, thinned residuals have the advantage that the
coordinates of the points are not transformed and, thus, the resulting
residuals may be easier to interpret. To obtain thinned residuals, each
point $(t_{i}, x_{i}, y_{i})$ is kept independently with probability
\[
\frac{b}{\hat{\lambda}(t_{i}, x_{i}, y_{i})},
\]
where
$b=\inf\{\hat{\lambda}(t,x,y)\dvtx (t,x,y) \in S\}$ is the infimum of the
estimated intensity over the entire observed space--time window, $S$.
The remaining points, called \textit{thinned residual points}, should be
homogeneous Poisson with rate $b$ if and only if the fitted model for
$\lambda$ is correct [\citet{Rick}]. For this method to have sufficient
power, several realizations of thinned residuals can be collected, each
realization being tested for uniformity using\vadjust{\goodbreak} the K-function, and then
all K-functions may be examined together to get the best overall
assessment of the model's fit.

When applied to the CSEP earthquake forecasts, $b$ tends to be so small
that thinning results in very few points (often zero) being retained.
One can instead obtain \textit{approximate thinned residuals} by forcing
the thinning procedure to keep, on average, a certain number, $k$, of
points by keeping each point with probability
\[
k\Big/\Biggl(\hat{\lambda}( t_{i}, x_{i}, y_{i})\sum_{i=1}^{N(S)}\hat{\lambda
}( t_{i}, x_{i}, y_{i})^{-1}\Biggr)
\]
as in \citet{Rick}.

%
\begin{figure}

\includegraphics{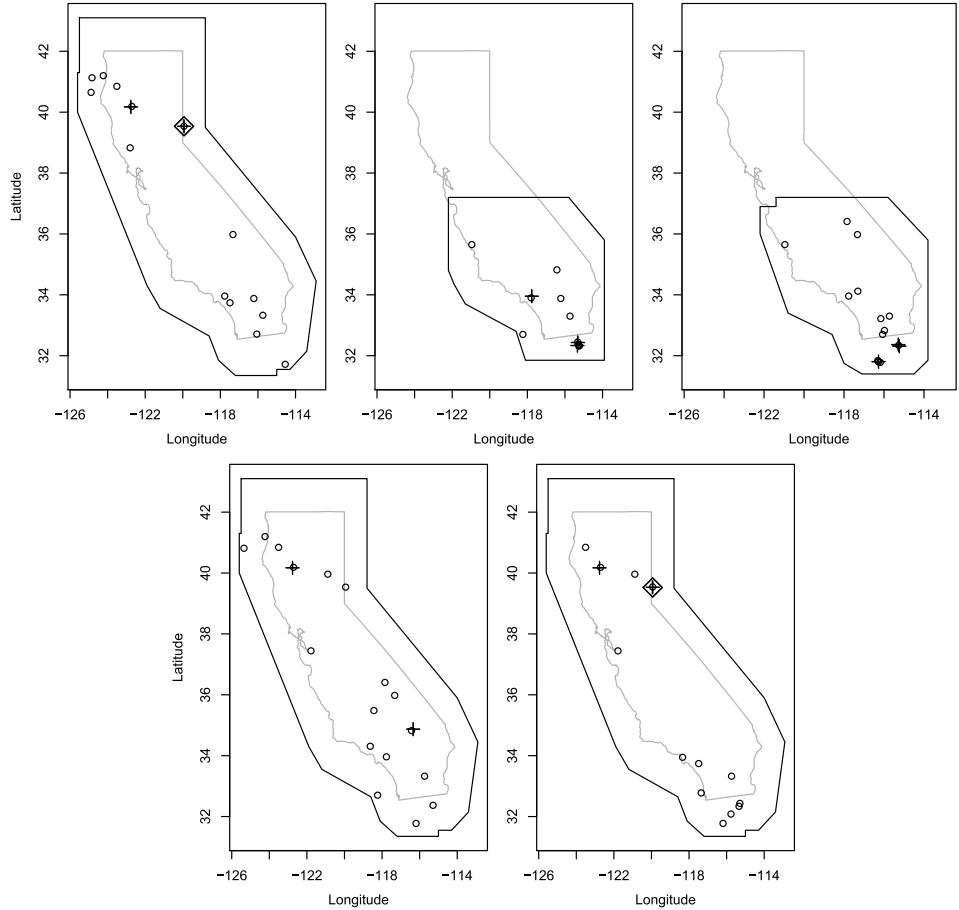}

\caption{One realization of thinned residuals for each of the five
models considered (nearby points are plotted with different symbols so
they can be differentiated). Top-left panel~\textup{(a)}:~model A ($k =
25$). Top-center panel \textup{(b)}: model B ($k = 15$). Top-right
panel~\textup{(c)}:~model~C ($k = 15$). Bottom-left panel \textup{(d)}:
ETAS ($k = 25$). Bottom-right panel~\textup{(e)}:~STEP ($k = 25$).}
\label{thinplotsall}
\end{figure}
%

%
\begin{figure}

\includegraphics{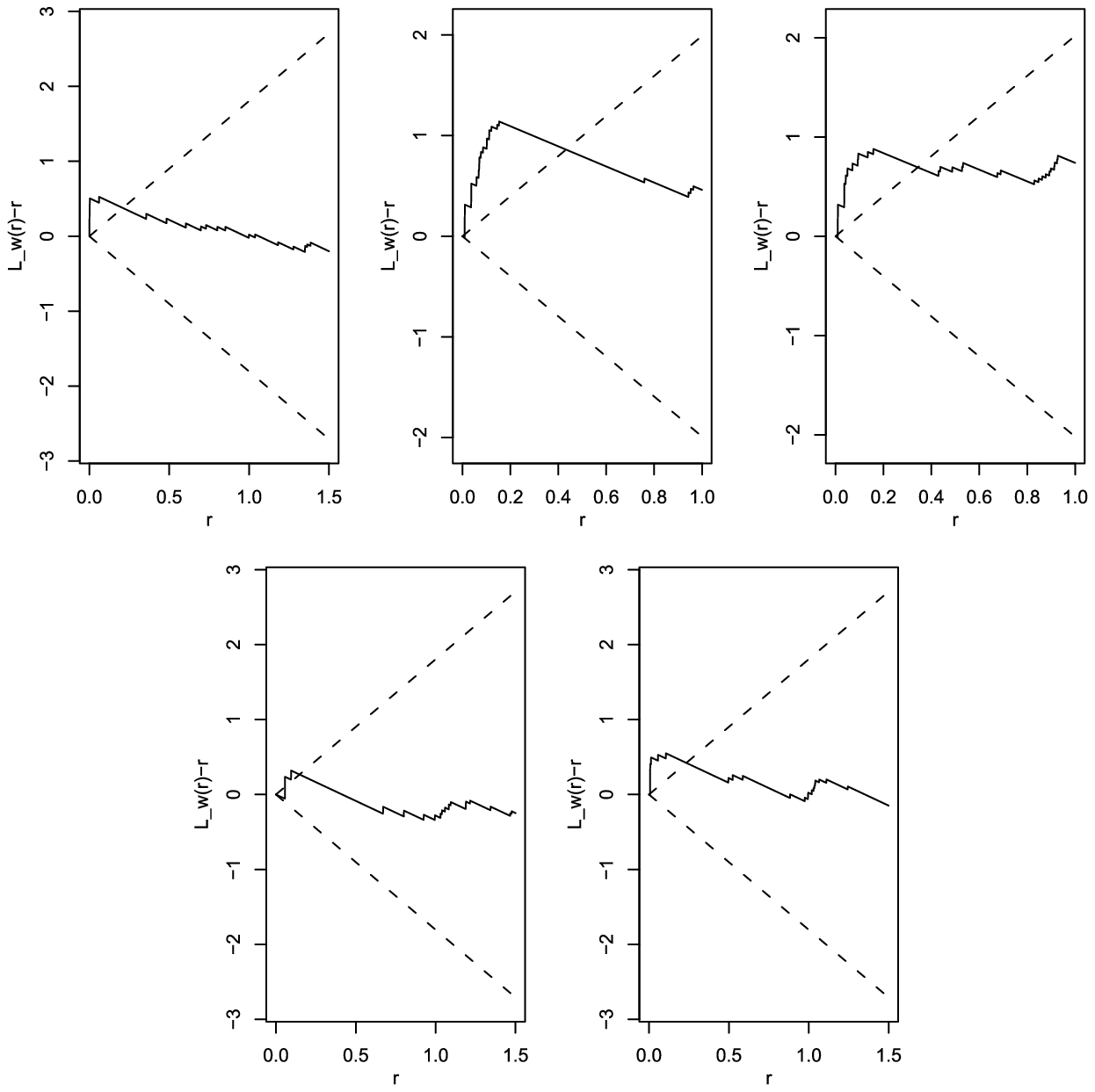}

\caption{Estimated centered weighted L-function (solid line) for one
realization of super-thinned residuals and 95\% bounds (dashed lines).
Top-left panel \textup{(a)}:\vspace*{1pt} model A
($\hat{\lambda}_{0} = 0.296$). Top-center
panel~\textup{(b)}: model B ($\hat{\lambda}_{0} = 0.406$).
Top-right panel \textup{(c)}:
model~C ($\hat{\lambda}_{0} = 0.394$). Bottom-left\vspace*{1pt} panel \textup{(d)}: ETAS
($\hat{\lambda}_{0} = 0.296$). Bottom-right panel \textup{(e)}: STEP
($\hat{\lambda}_{0} = 0.296$).}
\label{thinplotsallwk}
\end{figure}

Typical examples of approximate thinned residuals for the five models
we consider, using $k = 25, 15, 15, 25$ and $25$ for models A, B, C,
ETAS and STEP, respectively, are shown in Figure \ref{thinplotsall}.
Excessive clustering or inhibition in the residual process, compared
with what would be expected from a homogeneous Poisson process with
overall rate $k$, indicates lack of fit. To test the residuals for
homogeneity, one may apply the weighted K-function to the residuals,
with $\hat{\lambda}_{0}(\mathbf{x}_{i}) = k$ for all points
$\mathbf{x}_{i}$. This is equivalent to using the unweighted version of
the K-function on the residuals, except that here the overall rate is
$k$, whereas with the conventional unweighted K-function, the overall
rate is typically estimated as $N(S)/|S|$. The estimated centered
weighted L-functions for each model, along with the 95\%-confidence
bands based on \ref{wkbounds}, are shown in Figure
\ref{thinplotsallwk}. Models A and STEP most noticeably fail to thin
out the small cluster near the Peterson Mountain fault northwest of
Reno, Nevada, and another small cluster in northern California that
occurs approximately 35 kilometers south of the Battle Creek fault ($\mathrm{lon}
\approx122.7^{\circ}$W and $\mathrm{lat}\approx40.2^{\circ}$N). This
residual clustering is significant, as shown by the weighted
L-functions in Figures \ref{thinplotsallwk}(a) and (e). Model B has
trouble forecasting the Imperial cluster, as evidenced by the
significant clustering at distances up to 0.6$^{\circ}$. The residuals
for both models C and ETAS appear to be closer to uniformly distributed
throughout the space, though further investigation of several
realizations of thinned residuals reveals that model C has trouble
thinning out the Baja, California cluster, which leads to some
significant clustering in the residuals at very small distances.

\subsection{Superposition}

Superposition is a residual analysis technique similar to thinned
residuals, but instead of removing points, one simulates new points to
be added to the data and examines the result for uniformity. This
procedure was proposed by \citet{Bremaud}, but examples of its use have
been elusive. Points are simulated at each location $(t,x,y)$ according
to a~Cox process with intensity $c-\hat{\lambda}(t_{i},x_{i},y_{i})$,
where $c = \sup_{S}\{\hat{\lambda}(t,x,y)\}$. As with thinning and
rescaling, if the model for $\lambda$ is correct, the union of the
superimposed residuals and observed points will be homogeneous Poisson.
Any patterns of inhomogeneity in the residuals aid us in identifying
spots where the model fits poorly.

\begin{figure}

\includegraphics{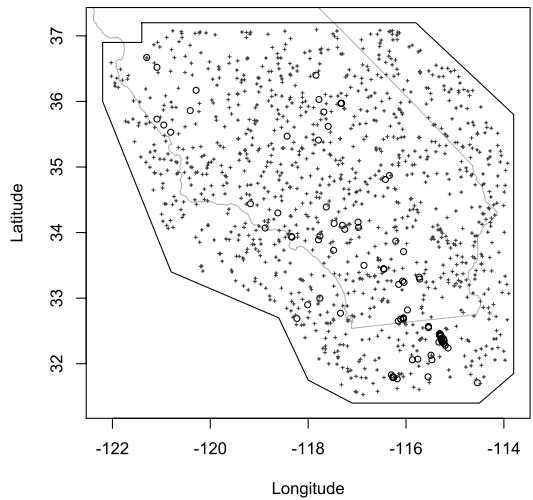}

\caption{Superposed residuals for model C. Simulated points make up
90.7\% of all points.}
\label{Shenxtsuper}
\end{figure}

Superposition helps solve one of the biggest disadvantages of thinned
residuals: the lack of information on the goodness of fit of the model
in locations where no events occur. However, if $c$ is large, then
there is a possibility that too many points will be simulated, meaning
that the behavior of the K-function will be primarily influenced by
simulated points rather than actually observed data points. For models
A and STEP, for example, simulated points comprise${}\geq{}$99\% of the
total points after superposition. For models C and ETAS, simulated
points comprise${}\geq{}$90\% of the superposed residual points. See
Figure \ref{Shenxtsuper} for an example of superposed residuals for
model C. Since the test for uniformity is based almost entirely on the
simulated points, which are by construction approximately homogeneous
for large~$c$, the test has low power for model evaluation in such
situations.

%
%

%
\begin{figure}

\includegraphics{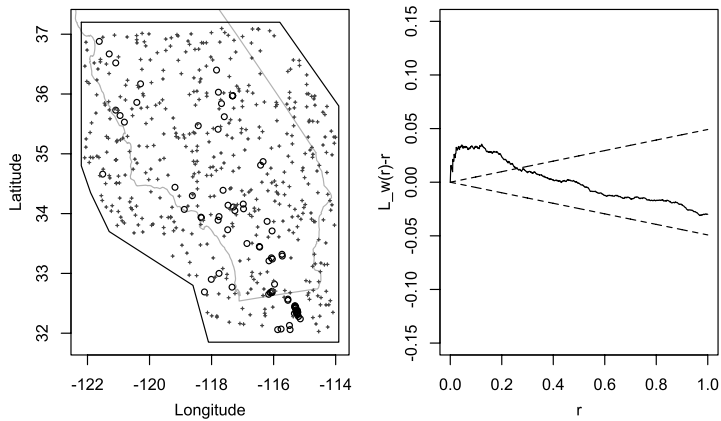}

\caption{Superposed residuals for model B. Left panel \textup{(a)}: one
realization of superposed residuals (circles${} = {}$observed earthquakes;
plus signs${} = {}$simulated points). Right panel~\textup{(b)}:~estimated centered
weighted L-function for superposed residuals (solid line) and
95\%-confidence bounds (dashed lines).}\label{Kaganxtsuperwk}
\end{figure}

A realization of superposed residuals for model B can be seen in Figure~\ref{Kaganxtsuperwk}, along with the corresponding centered weighted
L-function as a test for homogeneity of the residuals. 95\%-confidence
bands for the L-function are constructed under the null hypothesis $\hat
{\lambda}_{0}(\mathbf{x}_{i}) = c$ for all points $\mathbf{x}_{i}$. The
superposed residuals are significantly more clustered than would be
expected, up to distances of 0.4$^{\circ}$, or approximately 44.4 km.
This is likely the result of the underprediction of the seismicity rate
in the Imperial cluster.
One also observes significantly more inhibition in the superposed
residuals than would be expected at distances greater than~0.5$^{\circ
}$, or approximately 55.5 km. This inhibition can most likely be
attributed to the model's overprediction of the seismicity rate in
areas devoid of earthquakes, which can be seen in the portions of
Figure \ref{Kaganxtsuperwk}(a) in various regions lacking both
simulated and observed points.

\subsection{Super-thinning}

A more powerful approach than thinning or superposition individually is
a hybrid approach where one thins in areas of high intensity and
superposes simulated points in areas of low intensity, resulting in a
homogeneous point process if the model for $\lambda$ used in the
thinning and superposition is correct. The benefit of this method,
called super-thinning by \citet{Clements}, is that the user may specify
the overall rate of the resulting residual point process, $Z$, so that
it contains neither too few or too many points.

In super-thinning, one first keeps each observed point $(t,x,y)$ in the
catalog independently with probability $\min\{1, k/ \hat
\lambda(t,x,y)\}$ and subsequently superposes points
generated\vspace*{2pt}
according to a simulated Cox process with rate $\max\{0, k - \hat
\lambda(t,x,y)\}$. The result is a homogeneous Poisson process with
rate~$k$ if and only if the model $\hat\lambda$ for the conditional
intensity is correct [\citet{Clements}] and, hence, the resulting
super-thinned residuals can be assessed for homogeneity as a way of
evaluating the model. In particular, any clustering or inhibition in
the residual points indicates a lack of fit.

%
\begin{figure}

\includegraphics{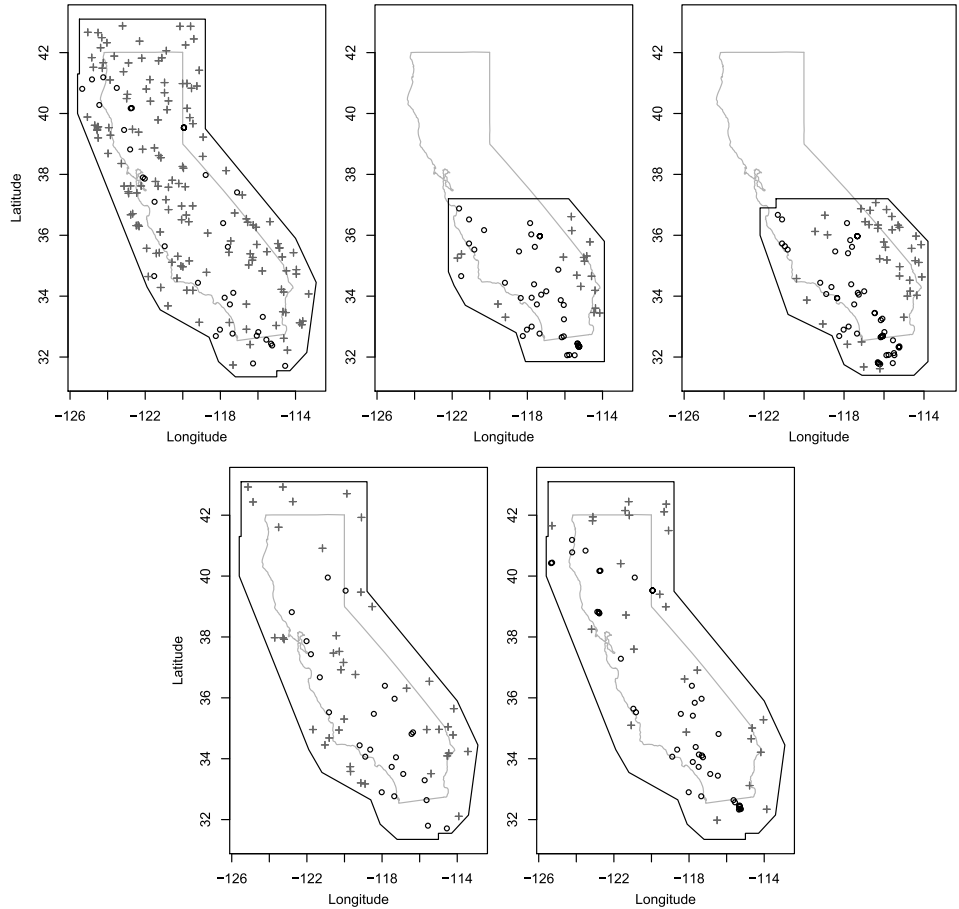}

\caption{One realization of super-thinned residuals for the five models
considered (circles${} = {}$observed earthquakes; plus signs${} =
{}$simulated points). Top-left panel~\textup{(a)}:~model~A ($k =
2.76$). Top-center panel~\textup{(b)}:~model~B ($k = 2.95$). Top-right
panel~\textup{(c)}:~model~C ($k = 2.73$). Bottom-left panel~\textup{(d)}:~ETAS ($k = 1.35$). Bottom-right panel~\textup{(e)}:~STEP
($k = 0.75$).}\label{allsuperthin}
\end{figure}
%

%
\begin{figure}

\includegraphics{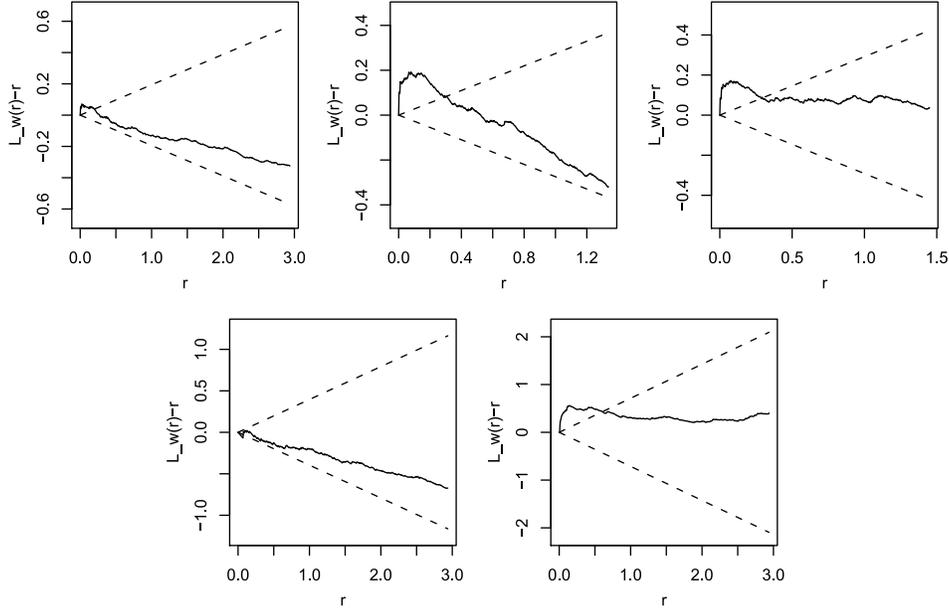}

\caption{Estimated centered weighted L-function (solid line) and 95\%
bands (dashed lines) for the super-thinned residuals in Figure
\protect\ref{allsuperthin}. Top-left panel \textup{(a)}: model A
($\hat{\lambda}_{0} = 2.76$). Top-center panel~\textup{(b)}: model B
($\hat{\lambda}_{0} = 2.95$). Top-right\vspace*{1pt} panel \textup{(c)}: model C
($\hat{\lambda}_{0} = 2.73$). Bottom-left panel \textup{(d)}: ETAS
($\hat{\lambda}_{0} = 1.35$). Bottom-right panel \textup{(e)}: STEP
($\hat{\lambda}_{0} = 0.75$).}\label{allsuperthinwk}
\end{figure}

In the application to earthquake forecasts, a natural choice for $k$ is
the total number of expected earthquakes according to each forecast.
Figure~\ref{allsuperthin} shows one realization of super-thinned
residuals for each model, and Figure~\ref{allsuperthinwk} shows the
estimated centered weighted L-functions for the corresponding
residuals, with $\hat{\lambda}_{0}(\mathbf{x}_{i}) = k$ for all points
$\mathbf{x}_{i}$, along with 95\%-confidence bands. Model A appears to
fit rather well overall, with some significant clustering in the
residuals at very small distances (from 0$^{\circ}$ to 0.1$^{\circ}$)
most likely attributable to the same small clusters that remained in
the thinned residuals. However, the L-function in Figure
\ref{allsuperthinwk}(a) reveals that there is somewhat more inhibition
in the residual process than we would expect. This is likely
attributable to model A's overprediction of the seismicity rate
especially in inter-fault zones. The super-thinned residuals for model
B contain a~few significant clusters (Imperial, Laguna Salada and
Panamint) and some slight inhibition due to overprediction of
seismicity in two regions devoid of any simulated points or retained
earthquakes: the San Diego-Imperial County areas and the Los
Angeles--San Bernardino areas. There is also significant clustering for
model~C up to distances of 0.2$^{\circ}$, particularly the Laguna
Salada, Baja and Panamint clusters. The ETAS residuals contain
significant clustering at distances up to 0.1$^{\circ}$, and this is
largely attributable to the Imperial cluster and to clusters in
Peterson Mountain and the Mt. Konocti area near Clearlake, California
at $\mathrm{lon} \approx122.1^{\circ}$W and $\mathrm{lat} \approx38.8^{\circ}$N. The
STEP residuals exhibit significant clustering at distances up to
0.4$^{\circ}$, with obvious clustering at Imperial, Peterson Mountain,
Battle Creek, Mt. Konocti and the Mendocino fault zone off the coast of
Northwest California.

\section{Summary}\label{sec7}

A litany of residual analysis methods for spatial point processes can
be implemented to assess the fit and reveal weaknesses in point process
models, and many of these methods provide more reliable estimates of
the overall fit and more detailed information than the L-test and
N-test. Rescaled residuals can assist in the evaluation of the overall
spatial fit, but are not easily interpretable due to the transformed
spatial window. Thinned residuals are much more easily interpretable,
but suffer from variability in the thinned residual point pattern and
low power if $b$ is too small. Superposition is similar to thinning in
that it also suffers from sampling variability and low power in the
case of a very large supremum of $\hat{\lambda}$.
Super-thinning appears to be a promising alternative, but, like
superposition, may have low power if the modeled intensity is extremely
volatile. Deviance residuals and weighted second-order statistics appear
to be quite powerful, especially for comparisons of competing models.

Clearly, the availability of a larger number of observed earthquakes in
the tests would lead to more detailed and more meaningful results, and
this suggests further decreasing the lower magnitude threshold.
However, considerations of catalog incompleteness at lower magnitudes,
as well as the fact that not all forecast models in the study are
capable of forecasting small events and their spatial-temporal
fluctuations, lead to limits on how low one may place the lower
magnitude threshold for the catalog. Indeed, lowering the threshold
requires stronger time-dependence of the models to account for the
short-term fluctuations of microseismicity. Due to these
considerations, CSEP sets the lower magnitude threshold in most cases
to 3.95 for the time-varying models like STEP and ETAS.

Overall, model A seems to be overpredicting seismicity at the time of
testing, but this may change once the forecast period is complete if
there is a greater amount of seismic activity. Models B and C appear to
be significantly underpredicting seismicity in many locations, and
unless the seismic activity in these regions slows down considerably,
these models will continue to underpredict for the remainder of the
forecast period. The spatial distribution of model A is quite accurate,
coupling forecasts of high conditional intensity in areas along active
faults with very low intensity forecasts in areas adjacent to these
faults which typically are devoid of earthquakes. Models B and C have
smooth spatial distributions yielding erroneously high forecasts at
distances far from any faults.

The question of what choice of $k$ is optimal in thinning or
super-thinning remains open for future research. Ideally, $k$ should be
chosen such that a~poorly fitting model is rejected with high
probability, while a ``correct'' or satisfactorily fitting model is
rejected with low probability (i.e., the Type~I error probability,
$\alpha$, is small). When thinning, we lose information when points are
removed, so we prefer to keep as many points as possible, while keeping
$\alpha$ low. With super-thinning, we would also ideally want to retain
many of the original points while simulating few points, so that any
assessment of the homogeneity of the residuals is not highly dependent
on the simulations. Simulation and theoretical studies are needed in
the future to compare the power of these goodness-of-fit measures under
various hypotheses.

\section*{Acknowledgments}
We thank Yan Kagan and Alejandro Veen for helpful comments, the
Advanced National Seismic System for the earthquake catalog data, and
the Collaboratory for the Study of Earthquake Predictability and the
Southern California Earthquake Center for supplying the earthquake
forecasts.


%

%
\printaddresses

\end{document}